\documentclass{desyproc}
\usepackage{calc}
\numberwithin{equation}{section}
\newlength{\figh}
\newcommand{\MS}{\ensuremath{\overline{\mathrm{MS}}}}
\newcommand{\1}[1]{\mathbf{1}^#1}
\newcommand{\2}[1]{\mathbf{2}^#1}
\newcommand{\3}[1]{\mathbf{3}^#1}
\newcommand{\F}[4]{\,_{#1}F_{#2}\left(\left.\begin{array}{c}#3\end{array}\right|#4\right)}

\begin{document}
\title{Effective weak Lagrangians in the Standard Model and $B$ decays}
\author{{\slshape Andrey Grozin}\\[1ex]
Budker Institute of Nuclear Physics, Novosibirsk, Russia}
\contribID{xy}
\desyproc{DESY-PROC-2013-03}
\acronym{HQ2013}
\maketitle

\begin{abstract}
Weak processes (e.g., $B$ decays) with characteristic energies $\ll M_W$
can be described by an effective theory
which does not contain $W$, $Z$ and other heavy particles (Higgs, $t$).
Its Lagrangian contains four-fermion interaction operators.
Essentially it is the theory proposed by Fermi
and improved by Feynman, Gell-Mann, Marshak, Sudarshan.
\end{abstract}

\section{Introduction}
\label{S:Intro}

We don't know \emph{all} physics up to \emph{infinitely high} energies
(or down to \emph{infinitely small} distances).
\emph{All} our theories are effective low-energy (or large-distance) theories
(except \emph{The Theory of Everything} if such a thing exists).
There is a high energy scale $M$ where an effective theory breaks down.
Its Lagrangian describes light particles ($m_i\ll M$)
and their interactions at low momenta ($p_i\ll M$).
In other words, it describes physics at large distances $\gg1/M$;
physics at small distances $\lesssim1/M$ produces local interactions of these
light fields.
The Lagrangian contains all possible operators (allowed by symmetries).
Coefficients of operators of dimension $n+4$ contain $1/M^n$.
If $M$ is much larger than energies we are interested in,
we can retain only renormalizable terms (dimension 4), and, maybe,
a power correction or two.

In order to describe weak processes with characteristic energies $\ll M_W$,
such as $b$ decays,
we can use an effective theory without $W^\pm$, $Z^0$, Higgs, $t$.
In these lectures we consider effective Lagrangians for some $b$ decay processes.
Coefficients of local interaction operators in this Lagrangian
are obtained by matching at $\mu\sim M_W$.
In order to calculate $b$ decays one needs to know these coefficients
at a much lower $\mu\sim m_b$.
They are obtained by solving renormalization group equations.
One needs to calculate the matrix of anomalous dimensions of the operators
entering the effective Lagrangian.

\begin{figure}[ht]
\begin{center}
\begin{picture}(112,72)
\put(56,36){\makebox(0,0){\includegraphics{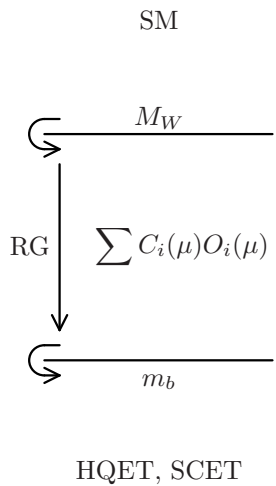}}}
\put(56,66){\makebox(0,0){SM}}
\put(59,36){\makebox(0,0){$\displaystyle\sum C_i(\mu) O_i(\mu)$}}
\put(56,6){\makebox(0,0){HQET, SCET}}
\put(56,53){\makebox(0,0){$M_W$}}
\put(56,18){\makebox(0,0){$m_b$}}
\put(39,36){\makebox(0,0){RG}}
\end{picture}
\end{center}
\caption{Hierarchy of effective theories.}
\label{F:hier}
\end{figure}

Of course, the knowledge of the Lagrangian is not sufficient.
In order to obtain full or differential decay rates into various channels,
we need to calculate these decay rates in the framework of the effective theory.
The largest energy scale in such calculations is $m_b$;
all information about physics at the scale $M_W$ is contained in the coefficients
of interaction operators in the effective Lagrangian.
For total decay rates into a channel with some flavor quantum numbers
if is sufficient to calculate the spectral density of the correlator
of the relevant interaction operators;
this is a single-scale problem with the scale $m_b$
(in some cases one has also to take $m_c\neq0$ into account).
For more detailed decay characteristics is is often useful to construct
further effective theories for energy scales $\ll m_b$ (HQET, SCET; Fig.~\ref{F:hier}).
One performs matching at $\mu\sim m_b$ to obtain coefficients
in such effective Lagrangians,
and then evolves them to lower $\mu$ using renormalization group.
We shall not discuss these questions here.

Effective Lagrangians for $B$ decays are discussed in great detail
in the excellent lectures by A.~Buras~\cite{B:98};
the reader is encouraged to use them for learning any information missing here.
References to the relevant papers can be found in~\cite{B:11}.
Here I don't cite original papers,
except a few ones which contain material directly used in these lectures.

The traditional Fermi constant $G$ is not used in these lectures,
because it is better to see the powers of $1/M_W$ and coupling constants explicitly.
We mainly work at the leading $1/M_W^2$ order,
see Sect.~\ref{S:l} for brief comments about $1/M_W^4$.
Powers of couplings depend on the process:
$g_2^2$ for ordinary weak decays, $g_2^2 e$ for $b\to s\gamma$,
$g_2^4$ for $B^0\leftrightarrow\bar{B}^0$ oscillations (Sect.~\ref{S:osc}).

The matrix $\gamma_5$ is not used.
Left fermion fields are used;
this is, of course, necessary, because left and right fields
interact differently in the Standard Model.
Some operators with left fields vanish at $d=4$
(and thus become evanescent, Sect.~\ref{S:eva});
this is the only role played by the index $L$~\cite{CMM:98}.

\section{$b\to c l^- \bar{\nu}_l$}
\label{S:l}

\begin{figure}[ht]
\begin{center}
\begin{picture}(80,40)
\put(18.5,21.5){\makebox(0,0){\includegraphics{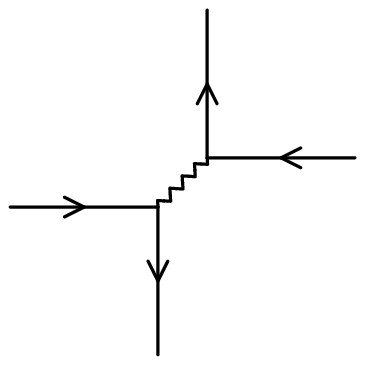}}}
\put(8.5,21.5){\makebox(0,0){$b$}}
\put(18.5,11.5){\makebox(0,0){$c$}}
\put(18.5,31.5){\makebox(0,0){$l^-$}}
\put(28.5,21.5){\makebox(0,0){$\nu_l$}}
\put(16.5,23.5){\makebox(0,0){$W$}}
\put(20.5,19.5){\makebox(0,0){$q$}}
\put(18.5,0){\makebox(0,0)[b]{a}}
\put(63.7,21.5){\makebox(0,0){\includegraphics{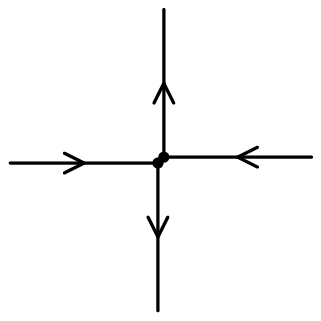}}}
\put(56.2,23.7){\makebox(0,0){$b$}}
\put(65.9,14){\makebox(0,0){$c$}}
\put(61.5,29){\makebox(0,0){$l^-$}}
\put(71.2,19.3){\makebox(0,0){$\nu_l$}}
\put(63.7,0){\makebox(0,0)[b]{b}}
\end{picture}
\end{center}
\caption{$b$ semileptonic decay in the full theory (a) and in the effective theory (b).}
\label{F:eft}
\end{figure}

The amplitude of the semileptonic decay $b\to c l^- \bar{\nu}_l$
in the Standard Model (Fig.~\ref{F:eft}a) is
\begin{equation}
M = \frac{g_2^2}{2} V_{cb} \frac{1}{M_W^2 - q^2}
(\bar{u}_{cL} \gamma^\alpha u_{bL})\,
(\bar{u}_{lL} \gamma_\alpha v_{\nu L})\,
\label{l:Mfull}
\end{equation}
where
\begin{equation*}
g_2 = \frac{e}{\sin\theta_W}
\end{equation*}
is the $SU(2)$ gauge coupling constant.
Expanding in $q^2/M_W^2\ll1$,
we have at the leading order
\begin{equation}
M = \frac{g_2^2}{2 M_W^2} V_{cb}
(\bar{u}_{cL} \gamma^\alpha u_{bL})\,
(\bar{u}_{lL} \gamma_\alpha v_{\nu L})\,.
\label{l:Meft}
\end{equation}
This amplitude can be reproduced from the effective Lagrangian
(Fig.~\ref{F:eft}b)
\begin{equation}
L = \frac{g_2^2}{2 M_W^2} V_{cb}
(\bar{c}_L \gamma^\alpha b_L)\,
(\bar{l}_L \gamma_\alpha \nu_L)\,.
\label{l:L}
\end{equation}

Now we shall discuss one-loop QCD renormalization of the operator
\begin{equation}
O_0 = (\bar{c}_{0L} \gamma^\alpha b_{0L})\,
(\bar{l}_L \gamma_\alpha \nu_L)
\label{l:O0}
\end{equation}
(we are not going to consider electroweak loop corrections;
therefore, the lepton fields don't renormalize).
This bare operator is related to the renormalized one as
\begin{equation}
O_0 = Z(\alpha_s(\mu)) O(\mu)\,,\qquad
O(\mu) = Z^{-1}(\alpha_s(\mu)) O_0
\label{l:Omu}
\end{equation}
in the \MS{} scheme.
In the matrix element of the bare operator
\begin{equation*}
{<}O_0{>} = Z {<}O{>}
\end{equation*}
$\alpha_s/\varepsilon$ term comes only from $Z$.
This matrix element is
\begin{equation}
\setlength{\figh}{(22.6mm-1ex)/2}
{<}O_0{>} = Z_q \left[
\raisebox{-\figh}{\includegraphics{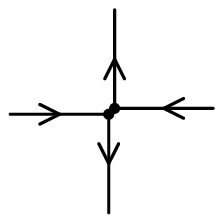}}
+ \raisebox{-\figh}{\includegraphics{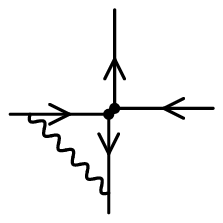}}
\right]\,,
\end{equation}
where $Z_q$ is the \MS{} quark field renormalization constant.
We need only the $1/\varepsilon$ term in the $\alpha_s$ correction.

\begin{figure}[ht]
\begin{center}
\begin{picture}(34.6,34.6)
\put(17.3,17.3){\makebox(0,0){\includegraphics{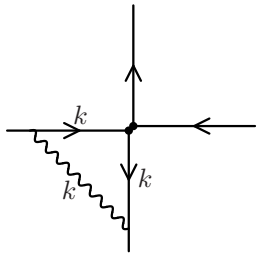}}}
\put(10.5,19){\makebox(0,0){$k$}}
\put(19,10.5){\makebox(0,0){$k$}}
\put(9,9){\makebox(0,0){$k$}}
\end{picture}
\end{center}
\caption{The one-loop correction to the vertex function of $O_0$.}
\label{F:V1a}
\end{figure}

The UV divergence of the vertex (Fig.~\ref{F:V1a})
does not depend on external momenta,
therefore we may set them to 0:
\begin{equation}
\begin{split}
\Lambda_1 &{} = -i C_F g_0^2 \int \frac{d^d k}{(2\pi)^d} \frac{1}{(k^2)^3}
\left( g_{\mu\nu} - \xi \frac{k_\mu k_\nu}{k^2} \right)
\gamma^\mu \rlap/k \gamma^\alpha \rlap/k \gamma^\nu \otimes \gamma_\alpha\\
&{} = -i C_F g_0^2 \int \frac{d^d k}{(2\pi)^d} \frac{1}{(k^2)^2}
\left[ \frac{1}{d} \gamma^\mu \gamma^\lambda \gamma^\alpha \gamma_\lambda \gamma_\mu - \xi \gamma^\alpha \right] \otimes \gamma_\alpha\\
&{} = -i C_F g_0^2 \int \frac{d^d k}{(2\pi)^d} \frac{1}{(k^2)^2}
\left[ \frac{(d-2)^2}{d} - \xi \right] \gamma^\alpha \otimes \gamma_\alpha\,
\end{split}
\label{l:V1}
\end{equation}
where $A\otimes B$ means $(\bar{u}_{cL} A u_{bL})\,(\bar{u}_{lL} B v_{\nu L})$,
and the gluon propagator is
\begin{equation*}
D_{\mu\nu}(k) = \frac{1}{k^2} \left[ g_{\mu\nu} - \xi \frac{k_\mu k_\nu}{k^2} \right]\,.
\end{equation*}
Of course, we need some IR regularization here,
e.g.\ a non-zero mass in the denominator:
\begin{equation}
\int \frac{d^d k}{(k^2)^2} \Rightarrow
\int \frac{d^d k}{(k^2-m^2)^2} \Rightarrow
\frac{i}{(4\pi)^2} \frac{1}{\varepsilon}
\label{l:IR}
\end{equation}
(this is a simplest example of infrared rearrangement).
Substituting the well-known one-loop $Z_q$
and keeping only $1/\varepsilon$ in the $\alpha_s$ correction,
we obtain
\begin{equation}
{<}O_0{>} = \left[ 1 - C_F \frac{\alpha_s}{4\pi\varepsilon} (1-\xi) \right]\,
\left[ 1 + C_F \frac{\alpha_s}{4\pi\varepsilon} (1-\xi) \right]
= 1\,.
\label{l:V1res}
\end{equation}
Hence $Z(\alpha_s) = 1$ --- the vector current does not renormalize.
This is true to all orders in $\alpha_s$ as follows from the Ward identity.
Note that we haven't used $\gamma_5$ in this calculation:
it is hidden in the index $L$ of the external fermion wave functions;
this implicitly means the anticommuting $\gamma_5$.
Hence the axial current with the anticommuting $\gamma_5$ does not renormalize too;
this is obvious --- we can always anticommute $\gamma_5$ out of the calculation.

It is not difficult to construct an effective Lagrangian
which reproduces results of the full theory expanded up to $1/M_W^4$.
The $b\to c l^- \bar{\nu}_l$ decay matrix element~(\ref{l:Mfull}
with this accuracy is
\begin{equation}
M = \frac{g_2^2}{2 M_W^2} V_{cb}
\left(1 + \frac{q^2}{M_W^2}\right)
(\bar{u}_{cL} \gamma^\alpha u_{bL})\,
(\bar{u}_{lL} \gamma_\alpha v_{\nu L})\,;
\label{l:Meft2}
\end{equation}
it follows from the effective Lagrangian
\begin{equation}
L = \frac{g_2^2}{2 M_W^2} V_{cb}
(\bar{c}_L \gamma^\alpha b_L)
\left(1 - \frac{\partial^2}{M_W^2}\right)
(\bar{l}_L \gamma_\alpha \nu_L)\,.
\label{l:L2}
\end{equation}
When calculating any process with the $1/M_W^4$ accuracy,
we can include in a diagram either a single $1/M_W^4$ vertex from the effective Lagrangian,
or up to two $1/M_W^2$ vertices.
We need to investigate renormalization of the dimension-8 operators
which appear in the $1/M_W^4$ term in the Lagrangian;
there is a finite number of such operators.
We also need to renormalize bilocal products of pairs of dimension-6 operators
which appear in the $1/M_W^2$ term in the Lagrangian.
In addition to renormalization of each operator,
local dimension-8 counterterms are needed.
In general,
at any order in $1/M_W^2$ a finite number of renormalization constant is needed,
and the theory retains its predictive power.

\section{$b \to c d \bar{u}$}
\label{S:u}

\subsection{Effective Lagrangian}
\label{S:uL}

In this Section we shall discuss the non-leptonic decay
where all four flavors are different
at the leading order in electroweak interaction.
Its full-theory matrix element at the $1/M_W^2$ level
is reproduced by the effective Lagrangian
\begin{equation}
L = \frac{g_2^2}{2 M_W^2} V_{cb} V_{ud}^*
(\bar{c}_L \gamma^\alpha b_L)\,
(\bar{d}_L \gamma_\alpha u_L)\,.
\label{u:L0}
\end{equation}

\begin{figure}[ht]
\begin{center}
\begin{picture}(80,40)
\put(18.5,21.5){\makebox(0,0){\includegraphics{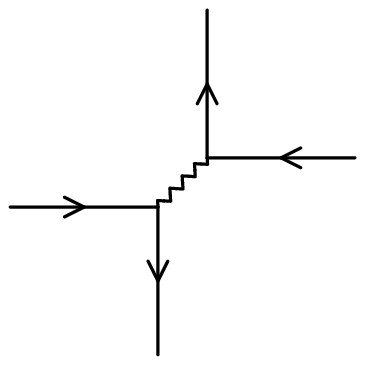}}}
\put(8.5,21.5){\makebox(0,0){$b$}}
\put(18.5,11.5){\makebox(0,0){$c$}}
\put(18.5,31.5){\makebox(0,0){$d$}}
\put(28.5,21.5){\makebox(0,0){$u$}}
\put(16.5,23.5){\makebox(0,0){$W$}}
\put(20.5,19.5){\makebox(0,0){$q$}}
\put(18.5,0){\makebox(0,0)[b]{a}}
\put(63.7,21.5){\makebox(0,0){\includegraphics{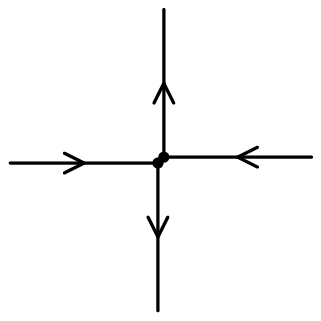}}}
\put(56.2,23.7){\makebox(0,0){$b$}}
\put(65.9,14){\makebox(0,0){$c$}}
\put(61.5,29){\makebox(0,0){$d$}}
\put(71.2,19.3){\makebox(0,0){$u$}}
\put(63.7,0){\makebox(0,0)[b]{b}}
\end{picture}
\end{center}
\caption{$b \to c d \bar{u}$ decay in the full theory (a) and in the effective theory (b).}
\label{F:cdu}
\end{figure}

We need to include a full set of operators closed under renormalization to the Lagrangian.
It consists of two operators
\begin{equation}
O_1 = (\bar{c}_{Li} \gamma^\alpha b_L^i)\,
(\bar{d}_{Lj} \gamma_\alpha u_L^j)\,,\qquad
O_2 = (\bar{c}_{Li} \gamma^\alpha b_L^j)\,
(\bar{d}_{Lj} \gamma_\alpha u_L^i)\,.
\label{u:O12}
\end{equation}
In $d=4$ we can use Fierz rearrangement%
\footnote{We know persons who became bosons.
Markus Fierz has become a verb: physicists say
``this can be proved by fierzing''
or ``let's fierz this product''.}
\begin{equation}
(\bar{\psi}_{1L} \gamma^\alpha \psi_{2L})\,(\bar{\psi}_{3L} \gamma_\alpha \psi_{4L})
= (\bar{\psi}_{3L} \gamma^\alpha \psi_{2L})\,(\bar{\psi}_{1L} \gamma_\alpha \psi_{4L})
\label{u:Fierz}
\end{equation}
to re-write these operators as
\begin{equation}
O_1 = (\bar{d}_{Lj} \gamma^\alpha b_L^i)\,
(\bar{c}_{Li} \gamma_\alpha u_L^j)\,,\qquad
O_2 = (\bar{d}_{Lj} \gamma^\alpha b_L^j)\,
(\bar{c}_{Li} \gamma_\alpha u_L^i)\,.
\label{u:O12f}
\end{equation}
Fierz rearrangement is especially simple~(\ref{u:Fierz})
in the case when all four wave functions are left:
there is exactly one structure possible in the right-hand side
($1\otimes1$, $\gamma_5\otimes\gamma_5$, $\sigma^{\alpha\beta}\otimes\sigma_{\alpha\beta}$ vanish;
$\gamma^\alpha\gamma_5\otimes\gamma_\alpha\gamma_5$ reduces to $\gamma^\alpha\otimes\gamma_\alpha$).

Sometimes the operator
\begin{equation}
O_2' = (\bar{c}_L t^a \gamma^\alpha b_L)\,
(\bar{d}_L t^a \gamma_\alpha u_L)
= T_F \left(O_2 - \frac{O_1}{N_c} \right)
\label{u:O2p}
\end{equation}
is used instead of $O_2$.
This relation follows from Cvitanovi\'c algorithm for $SU(N_c)$
\begin{align}
&(t^a)^i{}_j (t^a)^k{}_l
= T_F \left[ \delta^i_l \delta^k_j - \frac{1}{N_c} \delta^i_j \delta^k_l \right]\,,
\nonumber\\
&\setlength{\figh}{(12mm-1ex)/2}
\raisebox{-\figh}{\includegraphics{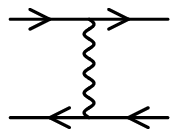}}
= T_F \left[ \raisebox{-\figh}{\includegraphics{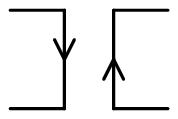}}
- \frac{1}{N_c} \raisebox{-\figh}{\includegraphics{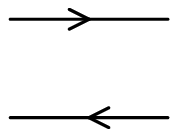}}
\right]
\label{u:Cvit}
\end{align}
(this is the color Fierz rearrangement).

The column vector of the bare operators $O_0$
is related to that of the renormalized operators $O(\mu)$ as
\begin{equation}
O_0 = Z(\alpha_s(\mu)) O(\mu)\,,\qquad
O(\mu) = Z^{-1}(\alpha_s(\mu)) O_0\,,
\label{u:ren}
\end{equation}
where $Z$ is the matrix of renormalization constants.
Differentiating this formula, we obtain the renormalization group equation
\begin{equation}
\frac{d O(\mu)}{d\log\mu} + \gamma(\alpha_s(\mu)) O(\mu) = 0\,,
\label{u:RG}
\end{equation}
where the anomalous dimension matrix is
\begin{equation}
\gamma = Z^{-1} \frac{d Z}{d\log\mu} = - \frac{d Z^{-1}}{d\log\mu} Z\,.
\label{u:gamma}
\end{equation}
The effective Lagrangian can be written via either bare or renormalized operators:
\begin{equation}
L = \frac{g_2^2}{2 M_W^2} V_{cb} V_{ud}^* c_0^T O_0
= \frac{g_2^2}{2 M_W^2} V_{cb} V_{ud}^* c^T(\mu) O(\mu)\,,
\label{u:L}
\end{equation}
where $c(\mu) = Z^T(\alpha_s(\mu)) c_0$ is the column vector
of Wilson coefficients.
It satisfies the RG equation
\begin{equation}
\frac{d c(\mu)}{d\log\mu} = \gamma^T(\alpha_s(\mu)) c(\mu)\,.
\label{u:RG1}
\end{equation}

Dividing~(\ref{u:RG1}) by the RG equation for $\alpha_s(\mu)$ we obtain
\begin{equation}
\frac{d c}{d\log\alpha_s} = - \frac{\gamma^T(\alpha_s)}{2\beta(\alpha_s)} c\,,
\label{u:RG2}
\end{equation}
where
\begin{equation*}
\beta(\alpha_s) = \beta_0 \frac{\alpha_s}{4\pi} + \cdots\,,\qquad
\gamma^T(\alpha_s) = \gamma^T_0 \frac{\alpha_s}{4\pi} + \cdots
\end{equation*}
At the leading (one-loop) order the solution is the matrix exponent
\begin{equation}
c(\mu) = \left(\frac{\alpha_s(\mu)}{\alpha_s(M_W)}\right)^{\textstyle-\frac{\gamma^T_0}{2\beta_0}} c(M_W)\,.
\label{u:c1}
\end{equation}
If eigenvectors $v_i$ of $\gamma^T_0$ ($\gamma^T_0 v_i = \lambda_i v_i$)
form a full basis%
\footnote{In some rare exceptional cases the Jordan form of $\gamma^T_0$ may contain blocks of sizes $>1$;
then the form of the solution is slightly different.}, then
\begin{equation}
c(\mu) = \sum A_i \left(\frac{\alpha_s(\mu)}{\alpha_s(M)}\right)^{\textstyle-\frac{\lambda_i}{2\beta_0}} v_i\,,
\label{u:c2}
\end{equation}
where $c(M_W) = \sum A_i v_i$.

The Wilson coefficients $c_i(\mu_0)$ at some scale $\mu_0$ are determined by matching ---
equating some $S$-matrix elements in the full theory (expanded in $p_i/M_W$) and in the effective theory.
It is most convenient to use $\mu_0\sim M_W$;
then $c_i(\mu_0)$ are given by perturbative series in $\alpha_s(\mu_0)$
containing no large logarithms.
They contain all the information about physics at the scale $M_W$
which is important for low-energy processes.
The Wilson coefficients $c_i(\mu)$ at low normalization scales $\mu$
are obtained by solving the RG equations.
The effective theory knows nothing about $M_W$;
the only information about it is contained in $c_i(\mu)$.
When the effective Lagrangian is applied to some physical process
with small momenta $p_i\ll M_W$,
it is most convenient to use $\mu$ of the order of the characteristic momenta:
then the results will contain no large logarithms.
This solution of the RG equation sums large logarithmic terms
in perturbation series.

\subsection{One-loop anomalous dimensions}
\label{S:u1}

The matrix element of the bare operator $O^0_1$
\begin{equation}
\setlength{\figh}{(22.6mm-1ex)/2}
\begin{split}
&{<}O_1^0{>} = Z_q^2 \Biggl[
\raisebox{-\figh}{\begin{picture}(22.6,22.6)
\put(11.3,11.3){\makebox(0,0){\includegraphics{grozin_andrey.fig04.eps}}}
\put(6,12){\makebox(0,0)[b]{$b$}}
\put(12,6){\makebox(0,0)[l]{$c$}}
\put(16.6,10.6){\makebox(0,0)[t]{$u$}}
\put(10.6,16.6){\makebox(0,0)[r]{$d$}}
\end{picture}}
+ \raisebox{-\figh}{\includegraphics{grozin_andrey.fig05.eps}}
+ \raisebox{-\figh}{\includegraphics{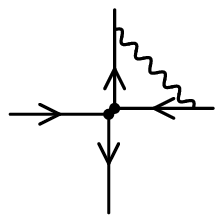}}\\
&{} + \raisebox{-\figh}{\includegraphics{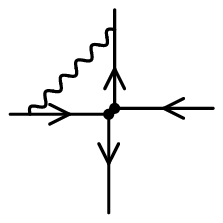}}
+ \raisebox{-\figh}{\includegraphics{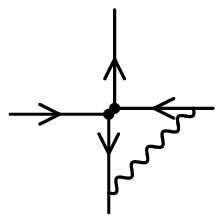}}
+ \raisebox{-\figh}{\includegraphics{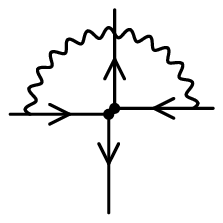}}
+ \raisebox{-\figh}{\includegraphics{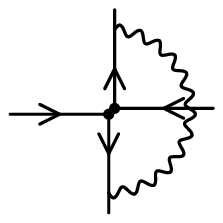}}
\Biggr]
\end{split}
\label{u:me}
\end{equation}
has two color structures
\begin{equation*}
T_1 = \delta_b^c \delta_u^d\,,\qquad
T_2 = \delta_b^d \delta_u^c
\end{equation*}
(the quark color indices coincide with the quark names).
The matrix element of $O^0_2$ can be obtained by simple substitutions
of the color structures.

\begin{figure}[ht]
\begin{center}
\begin{picture}(125,37.3)
\put(17.3,20.3){\makebox(0,0){\includegraphics{grozin_andrey.fig06.eps}}}
\put(10.5,22){\makebox(0,0){$k$}}
\put(19,13.5){\makebox(0,0){$k$}}
\put(9,12){\makebox(0,0){$k$}}
\put(17.3,0){\makebox(0,0)[b]{a}}
\put(62.5,20.3){\makebox(0,0){\includegraphics{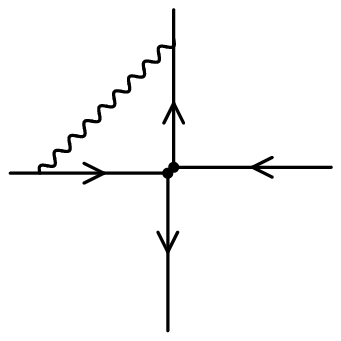}}}
\put(55.7,18){\makebox(0,0){$k$}}
\put(64.8,27.1){\makebox(0,0){$k$}}
\put(54.2,28.6){\makebox(0,0){$k$}}
\put(62.5,0){\makebox(0,0)[b]{b}}
\put(107.7,20.3){\makebox(0,0){\includegraphics{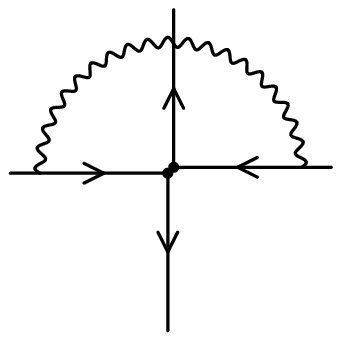}}}
\put(100.9,18){\makebox(0,0){$k$}}
\put(114.5,18.6){\makebox(0,0){$-k$}}
\put(105.7,35.3){\makebox(0,0){$k$}}
\put(107.7,0){\makebox(0,0)[b]{c}}
\end{picture}
\end{center}
\caption{One-loop $O^0_1$ vertex diagrams.}
\label{F:V1}
\end{figure}

The contribution of Fig.~\ref{F:V1}a differs from~(\ref{l:V1})
only by adding the color factor $T_1$:
\begin{equation}
\Lambda_1 = C_F T_1 \frac{\alpha_s}{4\pi\varepsilon} (1-\xi) \gamma^\alpha \otimes \gamma_\alpha\,.
\label{u:La1}
\end{equation}

Fig.~\ref{F:V1}b has the color structure $T_F(T_2-T_1/N_c)$~(\ref{u:Cvit}).
We only need the $1/\varepsilon$ UV divergence,
and hence we may do the $\gamma$-matrix algebra at $d=4$.
Fierz rearrangement makes this calculation identical to the previous one:
\begin{equation}
\Lambda_2 = T_F \left( T_2 - \frac{T_1}{N_c} \right) \frac{\alpha_s}{4\pi\varepsilon}
(1-\xi) \gamma^\alpha \otimes \gamma_\alpha\,.
\label{u:La2}
\end{equation}
We can also do this calculation explicitly:
\begin{equation}
\Lambda_2 = T_F \left( T_2 - \frac{T_1}{N_c} \right) \frac{\alpha_s}{4\pi\varepsilon}
\left[ \frac{1}{d} \gamma^\alpha \gamma^\lambda \gamma^\mu \otimes \gamma_\mu \gamma_\lambda \gamma_\alpha
- \xi \gamma^\alpha \otimes \gamma_\alpha \right]\,,
\label{u:La2a}
\end{equation}
where $\gamma^\mu\otimes\gamma_\mu$ comes from the gluon propagator,
and $\gamma^\lambda\otimes\gamma_\lambda$ from $\rlap/k\otimes\rlap/k$
after averaging over $k$ directions.
The $\gamma$-matrix structure appearing here
can be calculated at $d=4$ using Fierz rearrangement:
\begin{equation}
\gamma^\alpha \gamma^\lambda \gamma^\mu \otimes \gamma_\mu \gamma_\lambda \gamma_\alpha
= \gamma_\mu \gamma_\lambda \gamma^\alpha \gamma^\lambda \gamma^\mu \otimes \gamma_\alpha
= 4 \gamma^\alpha \otimes \gamma_\alpha\,,
\label{u1:fierz}
\end{equation}
and we again obtain~(\ref{u:La2}).

Fig.~\ref{F:V1}c also has the color structure $T_F(T_2-T_1/N_c)$.
It differs from Fig.~\ref{F:V1}b by the fact that
one $k$ is directed against the quark line (thus producing $-$),
and by the opposite order of $\gamma$-matrices on the second quark line:
\[
\Lambda_3 = - T_F \left( T_2 - \frac{T_1}{N_c} \right) \frac{\alpha_s}{4\pi\varepsilon}
\left[ \frac{1}{d} \gamma^\alpha \gamma^\lambda \gamma^\mu \otimes \gamma_\alpha \gamma_\lambda \gamma_\mu
- \xi \gamma^\alpha \otimes \gamma_\alpha \right]\,,
\]
cf.~(\ref{u:La2a}).
We can reduce this structure to the previous one
by anticommuting $\gamma$-matrices on the second line:
\[
\gamma_\alpha \gamma_\lambda \gamma_\mu = - \gamma_\mu \gamma_\lambda \gamma_\alpha
+ 2 \left( g_{\alpha\lambda} \gamma_\mu - g_{\alpha\mu} \gamma_\lambda + g_{\lambda\mu} \gamma_\alpha \right)\,,
\]
and hence
\begin{equation}
\gamma^\alpha \gamma^\lambda \gamma^\mu \otimes \gamma_\alpha \gamma_\lambda \gamma_\mu =
- \gamma^\alpha \gamma^\lambda \gamma^\mu \otimes \gamma_\mu \gamma_\lambda \gamma_\alpha
+ 2 (3d-2) \gamma^\alpha \otimes \gamma_\alpha\,.
\label{u:comm}
\end{equation}
Finally,
\begin{equation}
\Lambda_3 = - T_F \left( T_2 - \frac{T_1}{N_c} \right) \frac{\alpha_s}{4\pi\varepsilon}
(4-\xi) \gamma^\alpha \otimes \gamma_\alpha\,.
\label{u:La3}
\end{equation}

Adding mirror-symmetric diagrams and inserting the external leg renormalization $Z_q^2$,
we obtain the matrix element of the bare operator $O_1^0$:
\begin{equation}
\begin{split}
{<}O_1^0{>} ={}&
\left[ 1 - 2 C_F \frac{\alpha_s}{4\pi\varepsilon} (1-\xi) \right]\,
\biggl[ T_1\\
&{} + 2 C_F T_1 \frac{\alpha_s}{4\pi\varepsilon} (1-\xi)\\
&{} + 2 T_F \left( T_2 - \frac{T_1}{N_c} \right) \frac{\alpha_s}{4\pi\varepsilon} (1-\xi)\\
&{} - 2 T_F \left( T_2 - \frac{T_1}{N_c} \right) \frac{\alpha_s}{4\pi\varepsilon} (4-\xi)
\biggr] \gamma^\alpha \otimes \gamma_\alpha\\
={}& {<}O_1{>} - 6 T_F \frac{\alpha_s}{4\pi\varepsilon}
\left( {<}O_2{>} - \frac{{<}O_1{>}}{N_c} \right)\,.
\end{split}
\label{u:O1}
\end{equation}
It is gauge invariant, as expected.
In the case of the operator $O_2^0$,
Fig.~\ref{F:V1}b has the color structure $C_F T_2$,
and Fig.~\ref{F:V1}a, c --- $T_F(T_1-T_2/N_c)$~(\ref{u:Cvit}):
\begin{equation}
\begin{split}
{<}O_2^0{>} ={}&
\left[ 1 - 2 C_F \frac{\alpha_s}{4\pi\varepsilon} (1-\xi) \right]\,
\biggl[ T_2\\
&{} + 2 T_F \left( T_1 - \frac{T_2}{N_c} \right) \frac{\alpha_s}{4\pi\varepsilon} (1-\xi)\\
&{} + 2 C_F T_2 \frac{\alpha_s}{4\pi\varepsilon} (1-\xi)\\
&{} - 2 T_F \left( T_1 - \frac{T_2}{N_c} \right) \frac{\alpha_s}{4\pi\varepsilon} (4-\xi)
\biggr] \gamma^\alpha \otimes \gamma_\alpha\\
={}& {<}O_2{>} - 6 T_F \frac{\alpha_s}{4\pi\varepsilon}
\left( {<}O_1{>} - \frac{{<}O_2{>}}{N_c} \right)\,.
\end{split}
\label{u:O2}
\end{equation}

We arrive at the renormalization constant matrix
\begin{equation}
Z = 1 + 6 T_F \frac{\alpha_s}{4\pi\varepsilon}
\left(
\begin{array}{cc}
\frac{1}{N_c} & -1 \\
-1 & \frac{1}{N_c}
\end{array}
\right)
\label{u:Z1}
\end{equation}
at one loop.
In general, if
\[
Z = 1 + \frac{\alpha_s}{4\pi\varepsilon} z_1\,,
\]
then
\[\frac{d\,Z}{d\,\log\mu} = - 2 \varepsilon \frac{\alpha_s}{4\pi\varepsilon} z_1
= \gamma_0 \frac{\alpha_s}{4\pi}\,,
\]
and
\begin{equation}
\gamma_0 = - 2 z_1\,.
\label{u:gamma0}
\end{equation}
Therefore, in our case
\begin{equation}
\gamma_0 = - 12 T_F
\left(
\begin{array}{cc}
\frac{1}{N_c} & -1 \\
-1 & \frac{1}{N_c}
\end{array}
\right)\,.
\label{u:gamma1}
\end{equation}

It is easy to solve the eigenvalue problem $\gamma_0^T v_\pm = \lambda_\pm v_\pm$:
\begin{equation}
v_\pm = \left( \begin{array}{r} 1 \\ \pm1 \end{array} \right)\,,\qquad
\lambda_\pm = - 12 T_F \left( \frac{1}{N_c} \mp 1 \right)\,.
\label{u:vpm}
\end{equation}
Substituting the initial condition at $\mu=M_W$
\begin{equation}
c(M_W) = \left( \begin{array}{c} 1 \\ 0 \end{array} \right)
= \frac{1}{2} \left[ \left( \begin{array}{c} 1 \\ 1 \end{array} \right)
+ \left( \begin{array}{r} 1 \\ -1 \end{array} \right) \right]\,,
\label{u:MW}
\end{equation}
we obtain the running Wilson coefficients~(\ref{u:c2})
\begin{equation}
c(\mu) = \frac{1}{2} \left[ \left( \begin{array}{c} 1 \\ 1 \end{array} \right)
\left(\frac{\alpha_s(\mu)}{\alpha_s(M_W)}\right)^{\textstyle-\frac{\lambda_+}{2\beta_0}}
+ \left( \begin{array}{r} 1 \\ -1 \end{array} \right)
\left(\frac{\alpha_s(\mu)}{\alpha_s(M_W)}\right)^{\textstyle-\frac{\lambda_-}{2\beta_0}} \right]\,.
\label{u:cmu}
\end{equation}

Alternatively, one can introduce the operators
\begin{equation}
O_\pm = O_1 \pm O_2\,,
\label{u:Opm}
\end{equation}
so that
\[
L = c_+ O_+ + c_- O_-\,,\qquad
c_\pm = \frac{c_1 \pm c_2}{2}\,.
\]
With the one-loop accuracy, these operators renormalize independently:
\[
O^0_\pm = Z_\pm(\alpha_s(\mu)) O_\pm(\mu)\,.
\]
Substituting the initial conditions
\[
c_+(M_W) = c_-(M_W) = \frac{1}{2}\,,
\]
we obtain the one-loop running
\begin{equation}
c_\pm(\mu) = \frac{1}{2}
\left(\frac{\alpha_s(\mu)}{\alpha_s(M_W)}\right)^{\textstyle-\frac{\lambda_\pm}{2\beta_0}}\,.
\label{u:cpm}
\end{equation}
However, the operators $O_\pm$ do mix starting from two loops,
and therefore don't produce a great simplification.

\subsection{One-loop matching}
\label{S:um}

As already discussed, Wilson coefficients $c(\mu_0)$ ($\mu_0\sim M_W$)
are obtained by matching on-shell matrix elements in the full theory and the effective one.
Matching can be done at any on-shell momenta and quark masses;
it is most convenient to use the kinematic point where all $m_i=0$ and $p_i=0$.
The full-theory matrix elements should be expanded in $(m_i,p_i)/M_W$ to some order
for obtaining the coefficients in the effective Lagrangian
up to the corresponding order in $1/M_W$.
In particular, just setting all $m_i=0,$ $p_i=0$ produces the leading term
in this expansion, $1/M_W^2$.

With the one-loop accuracy the full-theory matrix element is
\begin{equation}
\setlength{\figh}{(26mm-1ex)/2}
\begin{split}
&(Z_q^{\text{os}})^2 \Biggl[
\raisebox{-\figh}{\includegraphics{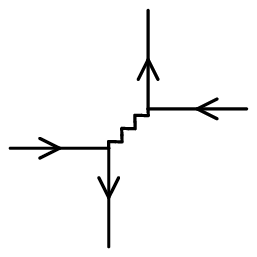}}
+ \raisebox{-\figh}{\includegraphics{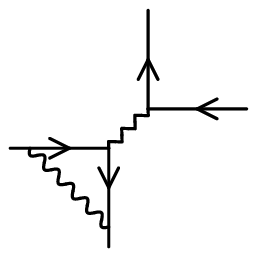}}
+ \raisebox{-\figh}{\includegraphics{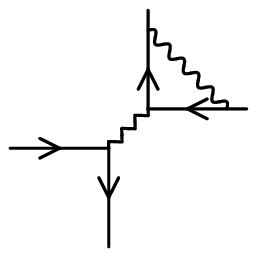}}\\
&{} + \raisebox{-\figh}{\includegraphics{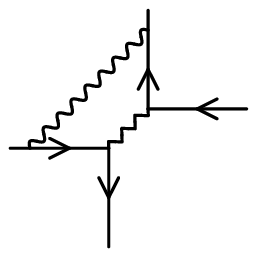}}
+ \raisebox{-\figh}{\includegraphics{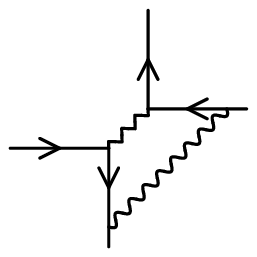}}
+ \raisebox{-\figh}{\includegraphics{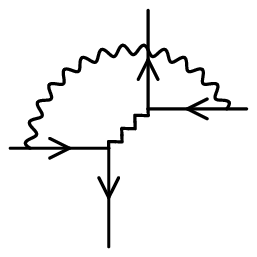}}
+ \raisebox{-\figh}{\includegraphics{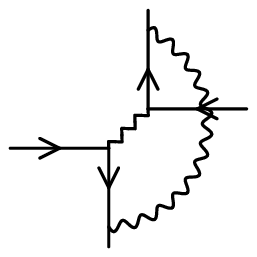}}
\Biggr]\,,
\end{split}
\label{Match:full}
\end{equation}
where $Z_q^{\text{os}}$ is the quark filed renormalization constant in the on-shell scheme
($Z_q=1$ if all $m_i=0$: loop corrections contain no scale).
The one-loop diagrams in the first line of this equation vanish:
they contain massless vacuum triangles with zero external momenta.

The effective-theory matrix element is given by the tree diagram Fig.~\ref{F:cdu}b
with the coupling constants $c_i^0$.
All loop corrections vanish because they are scale-free.
Note that the full-theory renormalized on shell matrix element
is UV finite but contains IR divergences.
The effective-theory one contains both UV and IR divergences which cancel each other
producing vanishing loop corrections.
IR divergences in the effective theory coincide with those in the full theory,
because the effective theory is designed to reproduce the small-momenta behavior of the full one.
Thus IR divergences cancel in the matching equation,
and $c_i^0$ contain UV $1/\varepsilon$ terms.
They are removed by renormalization when calculating $c_i(M_W)$.

\begin{figure}[ht]
\begin{center}
\begin{picture}(90,43)
\put(20,23){\makebox(0,0){\includegraphics{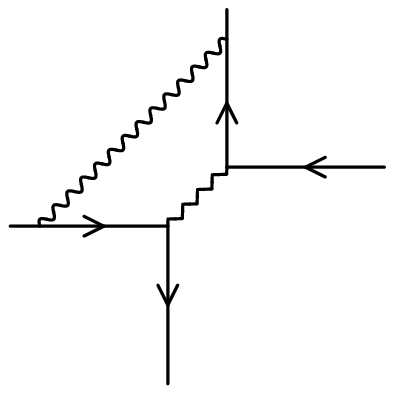}}}
\put(10.5,18){\makebox(0,0){$k$}}
\put(25,32.5){\makebox(0,0){$k$}}
\put(11.7,31.3){\makebox(0,0){$k$}}
\put(20,0){\makebox(0,0)[b]{a}}
\put(70,23){\makebox(0,0){\includegraphics{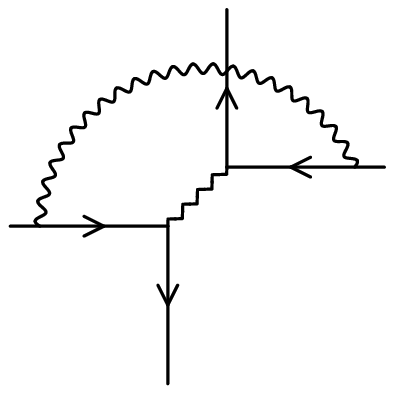}}}
\put(60.5,18){\makebox(0,0){$k$}}
\put(79.5,24){\makebox(0,0){$-k$}}
\put(70.5,38){\makebox(0,0){$k$}}
\put(70,0){\makebox(0,0)[b]{b}}
\end{picture}
\end{center}
\caption{One-loop full-theory diagrams.}
\label{F:match1}
\end{figure}

The diagram Fig.~\ref{F:match1}a is
\[
- i T_F \left( T_2 - \frac{T_1}{N_c} \right) g_0^2
\int \frac{d^d k}{(2\pi)^d}
\frac{\gamma^\alpha \rlap/k \gamma^\mu \otimes \gamma^\nu \rlap/k \gamma_\alpha}{(M_W^2 - k^2) (k^2)^3}
\left( g_{\mu\nu} - \xi \frac{k_\mu k_\nu}{k^2} \right)\,;
\]
after averaging over $k$ directions it becomes
\[
T_F \left( T_2 - \frac{T_1}{N_c} \right) \frac{g_0^2 M_W^{-2-2\varepsilon}}{(4\pi)^{d/2}} I
\left[ \frac{1}{d} \gamma^\alpha \gamma^\lambda \gamma^\mu \otimes \gamma_\mu \gamma_\lambda \gamma_\alpha
- \xi \gamma^\alpha \otimes \gamma_\alpha \right]\,,
\]
where the integral $I$ is
\[
\frac{1}{i\pi^{d/2}} \int \frac{d^d k}{(M_W^2 - k^2) (-k^2)^2} = I M_W^{d-6}
\]
(the power of $M_W$ is given by dimension counting).
Similarly, the diagram Fig.~\ref{F:match1}b is
\[
- T_F \left( T_2 - \frac{T_1}{N_c} \right) \frac{g_0^2 M_W^{-2-2\varepsilon}}{(4\pi)^{d/2}} I
\left[ \frac{1}{d} \gamma^\alpha \gamma^\lambda \gamma^\mu \otimes \gamma_\alpha \gamma_\lambda \gamma_\mu
- \xi \gamma^\alpha \otimes \gamma_\alpha \right]\,.
\]
It is easy to calculate $I$ using partial fractions:
\[
I = \frac{1}{i\pi^{d/2}} \int d^d k
\left[ \frac{1}{1-k^2} + \frac{1}{(-k^2)^2} - \frac{1}{-k^2} \right]
= \Gamma\left(1 - \frac{d}{2}\right)
\]
(we set $M_W=1$; integrals of powers of $-k^2$ vanish).
Adding mirror-symmetric diagrams, we obtain the full-theory matrix element
\[
\frac{1}{M_W^2}
\left[ T_1 \gamma^\alpha \otimes \gamma_\alpha
+ T_F \left( T_2 - \frac{T_1}{N_c} \right)
\frac{g_0^2 M_W^{-2\varepsilon}}{(4\pi)^{d/2}}
\frac{2}{d} \Gamma\left(1 - \frac{d}{2}\right)
\gamma^\alpha \gamma^\lambda \gamma^\mu \otimes
\left( \gamma_\mu \gamma_\lambda \gamma_\alpha
- \gamma_\alpha \gamma_\lambda \gamma_\mu \right)
\right]\,;
\]
using~(\ref{u:comm}) we arrive at
\begin{equation}
\begin{split}
&\frac{1}{M_W^2} \biggl\{ T_1 \gamma^\alpha \otimes \gamma_\alpha
- 12 T_F \left( T_2 - \frac{T_1}{N_c} \right)
\frac{g_0^2 M_W^{-2\varepsilon}}{(4\pi)^{d/2}}
\frac{1}{d} \Gamma\left(1 - \frac{d}{2}\right)\\
&\qquad{}\times\left[ (d-2) \gamma^\alpha \otimes \gamma_\alpha
- \frac{1}{3} \left( \gamma^\alpha \gamma^\beta \gamma^\gamma \otimes \gamma_\gamma \gamma_\beta \gamma_\alpha
- 4 \gamma^\alpha \otimes \gamma_\alpha \right) \right] \biggr\}\,.
\end{split}
\label{Match:fullres}
\end{equation}

In addition to
\[
\frac{1}{M_W^2} \left( c_1^0 T_1 + c_2^0 T_2 \right) \gamma^\alpha \otimes \gamma_\alpha
\]
(Fig.~\ref{F:cdu}b),
this result contains contributions of two bare evanescent operators $E^0_{1,2}$.
We shall see in Sect.~\ref{S:eva} that they are not zero;
however, they are equal to the renormalized $O_{1,2}(\mu)$ times factors containing $\alpha_s$,
and may be neglected with the present accuracy.
We obtain the bare Wilson coefficients
\begin{equation}
\begin{split}
&c_1^0 = 1 - 6 \frac{T_F}{N_c} \frac{g_0^2 M_W^{-2\varepsilon}}{(4\pi)^{d/2}} \Gamma(\varepsilon)
\left(1 + \frac{\varepsilon}{2} \right)\,,\\
&c_2^0 = 6 T_F \frac{g_0^2 M_W^{-2\varepsilon}}{(4\pi)^{d/2}} \Gamma(\varepsilon)
\left(1 + \frac{\varepsilon}{2} \right)\,.
\end{split}
\label{Match:c0}
\end{equation}
Using the renormalization constant matrix~(\ref{u:Z1})
we see that $1/\varepsilon$ UV divergences cancel in the renormalized Wilson coefficients:
\begin{equation}
\begin{split}
&c_1(\mu) = 1 - 12 \frac{T_F}{N_c} \frac{\alpha_s(\mu)}{4\pi}
\left( \log \frac{\mu}{M_W} + \frac{1}{4} \right)\,,\\
&c_2(\mu) = 12 T_F \frac{\alpha_s(\mu)}{4\pi}
\left( \log \frac{\mu}{M_W} + \frac{1}{4} \right)\,.
\end{split}
\label{Match:cmu}
\end{equation}
It is most convenient to perform matching at $\mu=M_Z$;
$c_i(M_Z)$ are given by series in $\alpha_s(M_Z)$ containing no logarithms:
\begin{equation}
\begin{split}
&c_1(M_W) = 1 - 3 \frac{T_F}{N_c} \frac{\alpha_s(M_W)}{4\pi}\,,\\
&c_2(M_W) = 3 T_F \frac{\alpha_s(M_W)}{4\pi}\,.
\end{split}
\label{Match:cM}
\end{equation}
They can be used as initial conditions for RG equations
to find $c_i(\mu)$ for $\mu\ll M_W$.

\subsection{Evanescent operators}
\label{S:eva}

In dimensional regularization we have to consider,
in addition to the physical operators
\begin{equation}
\begin{split}
&O^0_1 = (\bar{c}_{L0i} \gamma^\alpha b_{L0}^i)\,
(\bar{d}_{L0j} \gamma_\alpha u_{L0}^j)\,,\\
&O^0_2 = (\bar{c}_{L0i} \gamma^\alpha b_{L0}^j)\,
(\bar{d}_{L0j} \gamma_\alpha u_{L0}^i)\,,
\end{split}
\label{eva:O}
\end{equation}
also evanescent operators
\begin{equation}
\begin{split}
&E^0_1 = (\bar{c}_{L0i} \gamma^\alpha \gamma^\beta \gamma^\gamma b_{L0}^i)\,
(\bar{d}_{L0j} \gamma_\gamma \gamma_\beta \gamma_\alpha u_{L0}^j)
- 4 O^0_1\,,\\
&E^0_2 = (\bar{c}_{L0i} \gamma^\alpha \gamma^\beta \gamma^\gamma b_{L0}^j)\,
(\bar{d}_{L0j} \gamma_\gamma \gamma_\beta \gamma_\alpha u_{L0}^i)
- 4 O^0_2\,.
\end{split}
\label{eva:E}
\end{equation}
At $d=4$, Fierz rearrangement
\begin{align*}
&(\bar{c}_{L0i} \gamma^\alpha \gamma^\beta \gamma^\gamma b_{L0}^i)\,
(\bar{d}_{L0j} \gamma_\gamma \gamma_\beta \gamma_\alpha u_{L0}^j)
= (\bar{d}_{L0j} \gamma_\gamma \gamma_\beta \gamma^\alpha \gamma^\beta \gamma^\gamma b_{L0}^i)\,
(\bar{c}_{L0i} \gamma_\alpha u_{L0}^j)\\
&{} = 4 (\bar{d}_{L0j} \gamma^\alpha b_{L0}^i)\,(\bar{c}_{L0i} \gamma_\alpha u_{L0}^j)
= 4 (\bar{c}_{L0i} \gamma^\alpha b_{L0}^i)\,(\bar{d}_{L0j}  \gamma_\alpha u_{L0}^j)
\end{align*}
states that they vanish.
However, these bare operators exist at $d\ne4$.
If we use the standard \MS{} renormalization prescription,
we'll see that the renormalized operators $E_{1,2}(\mu)$ also don't vanish.
This is not what we want.
Therefore we have to modify the \MS{} prescription to ensure vanishing
of renormalized evanescent operators~\cite{DG:91}.

\begin{sloppypar}
Recall (Sect.~\ref{S:u1}) that the renormalized matrix element of the bare operator
$(\bar{c}_{L0i} \Gamma b_{L0}^i)\,(\bar{d}_{L0j} \bar{\Gamma} u_{L0}^j)$ up to one loop,
\[
\setlength{\figh}{(22.6mm-1ex)/2}
Z_q^2 \Biggl[
\raisebox{-\figh}{\includegraphics{grozin_andrey.fig04.eps}}
+ 2 \raisebox{-\figh}{\begin{picture}(22.6,2.6)
\put(11.3,11.3){\makebox(0,0){\includegraphics{grozin_andrey.fig05.eps}}}
\put(3,13){\makebox(0,0){$\strut\mu$}}
\put(7,13){\makebox(0,0){$\strut\lambda$}}
\put(12.5,3){\makebox(0,0){$\strut\mu$}}
\put(12.5,7){\makebox(0,0){$\strut\lambda$}}
\end{picture}}
+ 2 \raisebox{-\figh}{\begin{picture}(22.6,2.6)
\put(11.3,11.3){\makebox(0,0){\includegraphics{grozin_andrey.fig11.eps}}}
\put(3,9){\makebox(0,0){$\strut\mu$}}
\put(7,9){\makebox(0,0){$\strut\lambda$}}
\put(13.1,19.6){\makebox(0,0){$\strut\mu$}}
\put(13.1,15){\makebox(0,0){$\strut\lambda$}}
\end{picture}}
+ 2 \raisebox{-\figh}{\begin{picture}(22.6,2.6)
\put(11.3,11.3){\makebox(0,0){\includegraphics{grozin_andrey.fig13.eps}}}
\put(3,9){\makebox(0,0){$\strut\mu$}}
\put(7,9){\makebox(0,0){$\strut\lambda$}}
\put(19.6,9.6){\makebox(0,0){$\strut\mu$}}
\put(15.6,9.6){\makebox(0,0){$\strut\lambda$}}
\end{picture}}
\Biggr]\,,
\]
is
\begin{equation}
\begin{split}
&\left( 1 - 2 C_F \frac{\alpha_s}{4\pi\varepsilon} \right) T_1 \Gamma \otimes \bar{\Gamma}
+ 2 C_F T_1 \frac{\alpha_s}{4\pi\varepsilon} \frac{1}{d} \gamma^\mu \gamma^\lambda \Gamma \gamma_\lambda \gamma_\mu \otimes \bar{\Gamma}\\
&{} + 2 T_F \left( T_2 - \frac{T_1}{N_c} \right) \frac{\alpha_s}{4\pi\varepsilon} \frac{1}{d} \Gamma \gamma^\lambda \gamma^\mu \otimes \gamma_\mu \gamma_\lambda \bar{\Gamma}
- 2 T_F \left( T_2 - \frac{T_1}{N_c} \right) \frac{\alpha_s}{4\pi\varepsilon} \frac{1}{d} \Gamma \gamma^\lambda \gamma^\mu \otimes \bar{\Gamma}\gamma_\lambda \gamma_\mu\,.
\end{split}
\label{eva:M1}
\end{equation}
For the operator
$(\bar{c}_{L0i} \Gamma b_{L0}^j)\,(\bar{d}_{L0j} \bar{\Gamma} u_{L0}^i)$
we have to adjust the color structures:
\begin{equation}
\begin{split}
&\left( 1 - 2 C_F \frac{\alpha_s}{4\pi\varepsilon} \right) T_2 \Gamma \otimes \bar{\Gamma}
+ 2 C_F T_2 \frac{\alpha_s}{4\pi\varepsilon} \frac{1}{d} \Gamma \gamma^\lambda \gamma^\mu \otimes \gamma_\mu \gamma_\lambda \bar{\Gamma}\\
&{} + 2 T_F \left( T_1 - \frac{T_2}{N_c} \right) \frac{\alpha_s}{4\pi\varepsilon} \frac{1}{d} \gamma^\mu \gamma^\lambda \Gamma \gamma_\lambda \gamma_\mu \otimes \bar{\Gamma}
- 2 T_F \left( T_1 - \frac{T_2}{N_c} \right) \frac{\alpha_s}{4\pi\varepsilon} \frac{1}{d} \Gamma \gamma^\lambda \gamma^\mu \otimes \bar{\Gamma}\gamma_\lambda \gamma_\mu\,.
\end{split}
\label{eva:M2}
\end{equation}
We obtain the matrix elements
\begin{equation}
\begin{split}
{<}O^0_1{>} ={}& T_1 \hat{O}
+ T_F \left( T_2 - \frac{T_1}{N_c} \right) \frac{\alpha_s}{4\pi\varepsilon} \left( - 6 \hat{O} + \hat{E} \right)\,,\\
{<}O^0_2{>} ={}& T_2 \hat{O} + C_F T_2 \frac{\alpha_s}{4\pi\varepsilon} \frac{1}{2} \hat{E}
+ T_F \left( T_1 - \frac{T_2}{N_c} \right) \frac{\alpha_s}{4\pi\varepsilon} \left( - 6 \hat{O} + \frac{1}{2} \hat{E} \right)\,,\\
{<}E^0_1{>} ={}& T_1 \hat{E} - C_F T_1 \frac{\alpha_s}{4\pi\varepsilon} 48 \varepsilon \hat{O}
+ T_F \left( T_2 - \frac{T_1}{N_c} \right) \frac{\alpha_s}{4\pi\varepsilon}
\left( - 48 \varepsilon \hat{O} - 14 \hat{E} + \hat{F} \right)\,,\\
{<}E^0_2{>} ={}& T_2 \hat{E}
+ T_F \left( T_1 - \frac{T_2}{N_c} \right) \frac{\alpha_s}{4\pi\varepsilon}
\left( 96 \varepsilon \hat{O} - 10 \hat{E} + \frac{1}{2} \hat{F} \right)\,,
\end{split}
\label{eva:M}
\end{equation}
where
\begin{align*}
&\hat{O} = \gamma^\alpha \otimes \gamma_\alpha\,,\\
&\hat{E} = \gamma^\alpha \gamma^\beta \gamma^\gamma \otimes \gamma_\gamma \gamma_\beta \gamma_\alpha
- 4 \hat{O}\,,\\
&\hat{F} = \gamma^\alpha \gamma^\beta \gamma^\gamma \gamma^\delta \gamma^\varepsilon \otimes \gamma_\varepsilon \gamma_\delta \gamma_\gamma \gamma_\beta \gamma_\alpha
- 16 \hat{O}
\end{align*}
are the $\gamma$-matrix structures of the physical operators $O_i$,
the evanescent operators $E_i$,
and the further evanescent operators
\begin{equation}
\begin{split}
&F^0_1 = (\bar{c}_{L0i} \gamma^\alpha \gamma^\beta \gamma^\gamma \gamma^\delta \gamma^\varepsilon b_{L0}^i)\,
(\bar{d}_{L0j} \gamma_\varepsilon \gamma_\delta \gamma_\gamma \gamma_\beta \gamma_\alpha u_{L0}^j)
- 16 O^0_1\,,\\
&F^0_2 = (\bar{c}_{L0i} \gamma^\alpha \gamma^\beta \gamma^\gamma \gamma^\delta \gamma^\varepsilon b_{L0}^j)\,
(\bar{d}_{L0j} \gamma_\varepsilon \gamma_\delta \gamma_\gamma \gamma_\beta \gamma_\alpha u_{L0}^i)
- 16 O^0_2\,.
\end{split}
\label{eva:F}
\end{equation}
which we need to introduce for calculating one-loop corrections to $E_i$.
\end{sloppypar}

We want renormalized evanescent operators to vanish:
\begin{equation}
\left(\begin{array}{c}O_0\\E_0\end{array}\right)
= Z(\alpha_s(\mu))
\left(\begin{array}{c}O(\mu)\\0\end{array}\right)\,,
\qquad
E(\mu) = 0\,.
\label{eva:Zdef}
\end{equation}
This vanishing should not be spoiled by the RG evolution.
Therefore, the anomalous dimension matrix should have the structure
\begin{align}
&\gamma(\alpha_s) = \left(
\begin{array}{cc}
\gamma_{OO} & \gamma_{OE} \\
0          & \gamma_{EE}
\end{array}
\right)\,,
\label{eva:gamma}\\
&\frac{d}{d\,\log\mu}
\left(\begin{array}{c}O(\mu)\\0\end{array}\right)
+ \left(
\begin{array}{cc}
\gamma_{OO} & \gamma_{OE} \\
0          & \gamma_{EE}
\end{array}
\right)
\left(\begin{array}{c}O(\mu)\\0\end{array}\right)\,,
\label{eva:RGO}
\end{align}
The evolution of the physical operators is not affected by evanescent ones:
\begin{equation}
\frac{d\,O(\mu)}{d\,\log\mu} + \gamma_{OO}(\alpha_s(\mu)) O(\mu) = 0\,.
\label{eva:RGO1}
\end{equation}
The RG evolution of the Wilson coefficients
of the physical ($c_O$) and evanescent ($c_E$) operators
is given by
\[
\frac{d}{d\,\log\mu}
\left(\begin{array}{c}c_O(\mu)\\c_E(\mu)\end{array}\right)
= \left(
\begin{array}{cc}
\gamma_{OO}^T & 0 \\
\gamma_{OE}^T & \gamma_{EE}^T
\end{array}
\right)
\left(\begin{array}{c}c_O(\mu)\\c_E(\mu)\end{array}\right)\,,
\]
or
\begin{equation}
\begin{split}
&\frac{d\,c_O(\mu)}{d\,\log\mu} = \gamma_{OO}^T c_O(\mu)\,,\\
&\frac{d\,c_E(\mu)}{d\,\log\mu} = \gamma_{OE}^T c_O(\mu) + \gamma_{EE}^T c_E(\mu)\,.
\end{split}
\label{eva:RGc}
\end{equation}
The evolution of $c_O(\mu)$ does not involve $c_E(\mu)$;
$c_E(\mu)\ne0$, but they are irrelevant because they are multiplied by $E(\mu)=0$.

Now let's have a close look at the one-loop matrix elements~(\ref{eva:M}).
We see that the matrix elements of the bare evanescent operators $E_i$
contain terms with the physical $\gamma$-matrix structure $\hat{O}$
finite at $\varepsilon\to0$!
When we start from an evanescent $\gamma$-matrix structure (such as $\hat{E}$),
which is 0 at $d=4$,
multiply it by some additional $\gamma$-matrices from a one-loop diagram,
and extract a physical $\gamma$-matrix structure (such as $\hat{O}$),
the coefficient must be proportional to $\varepsilon$.
However, when it is multiplied by $1/\varepsilon$ from the UV divergence of the loop integral,
the result is a finite contribution.
This UV divergence does not depend on external momenta and masses,
hence this physical term in the matrix element is similarly universal.
Therefore we can use the one-loop renormalization constant of the form
\begin{equation}
Z = 1 + \left(
\begin{array}{cc}
b            & c \\
a\varepsilon & d
\end{array}
\right)
\frac{\alpha_s}{4\pi\varepsilon}\,,
\label{eva:Z1str}
\end{equation}
so that
\begin{equation}
{<}E(\mu){>} = 0\,,\qquad
{<}E_0{>} = a {<}O(\mu){>} \frac{\alpha_s}{4\pi}\,.
\label{eva:E0mu}
\end{equation}
In other words, the renormalization constant is no longer minimal:
\begin{equation}
\begin{split}
&Z = 1 + \left( Z_{10} + \frac{Z_{11}}{\varepsilon} \right) \frac{\alpha_s}{4\pi}\,,\\
&Z_{10} = \left(
\begin{array}{cc}
0 & 0 \\
a & 0
\end{array}
\right)\,,\qquad
Z_{11} = \left(
\begin{array}{cc}
b & c \\
0 & d
\end{array}
\right)\,.
\end{split}
\label{eva:Z1}
\end{equation}
The anomalous dimension~(\ref{u:gamma}) is
\begin{equation}
\gamma = \gamma_0 \frac{\alpha_s}{4\pi}\,,\qquad
\gamma_0 = - 2 Z_{11}\,;
\label{eva:gamma1}
\end{equation}
it has the required structure~(\ref{eva:gamma}).
When calculating the one-loop anomalous dimension,
it is safe to forget about evanescent operators
(as we did in Sect.~\ref{S:u1}).

Now let's discuss renormalization at two loops.
We need a non-minimal renormalization matrix
\begin{equation}
Z(\alpha_s) = 1
+ \left( Z_{10} + \frac{Z_{11}}{\varepsilon} \right) \frac{\alpha_s}{4\pi}
+ \left( Z_{20} + \frac{Z_{21}}{\varepsilon} + \frac{Z_{22}}{\varepsilon^2} \right)
\left(\frac{\alpha_s}{4\pi}\right)^2\,.
\label{eva:Z2}
\end{equation}
The anomalous dimension matrix must be finite at $\varepsilon\to0$;
from this requirement we obtain
\begin{equation}
Z_{22} = \frac{1}{2} Z_{11} (Z_{11} -\beta_0)
= \frac{1}{2} \left(
\begin{array}{cc}
b (b-\beta_0) & b c + c d - \beta_0 c \\
0             & d (d-\beta_0)
\end{array}
\right)\,.
\label{eva:cons1}
\end{equation}
As usual, $1/\varepsilon^2$ terms in the two-loop $Z$ are not independent ---
they are given by products of one-loop terms.
The lower left corner is 0:
$\varepsilon$ from $\gamma$-matrix algebra moves this term to $Z_{21}$, see below.
Supposing that the self-consistency condition~(\ref{eva:cons1}) is satisfied,
the anomalous dimension matrix~(\ref{u:gamma}) is
\begin{equation}
\gamma(\alpha_s) = \gamma_0 \frac{\alpha_s}{4\pi} + \gamma_1 \left(\frac{\alpha_s}{4\pi}\right)^2\,,\quad
\gamma_0 = - 2 Z_{11}\,,\quad
\gamma_1 = - 2 (2 Z_{21} - Z_{10} Z_{11} - Z_{11} Z_{10} + \beta_0 Z_{10})\,.
\label{eva:gamma2}
\end{equation}
Let
\begin{equation}
Z_{21} = \left(
\begin{array}{cc}
e & f\\
g & h
\end{array}
\right)\,;
\label{eva:z21}
\end{equation}
$g$  is the $1/\varepsilon^2$ divergences of the two-loop integral
(which does not depend on external momenta) times $\varepsilon$ from $\gamma$-matrix algebra.
The lower left corner of $\gamma_1$ must vanish;
this gives the second self-consistency condition
\begin{equation}
g = \frac{1}{2} (ab + da - \beta_0 a)\,.
\label{eva:g}
\end{equation}
This contribution of $1/\varepsilon^2$ two-loop divergences
is also given by products of one-loop terms.

What we are really interested in is the upper left corner $\gamma_{OO}$
which determines the evolution of physical operators and Wilson coefficients.
With the two-loop accuracy
\begin{equation}
\gamma_{OO} = - 2 b \frac{\alpha_s}{4\pi}
- 2 (2 e + ca) \left(\frac{\alpha_s}{4\pi}\right)^2\,.
\label{eva:gammaoo}
\end{equation}
To calculate it correctly,
we need not only $e$ ---
the $1/\varepsilon$ part of two-loop diagrams with the insertion of a physical operator,
but also $a$ and $c$ --- one-loop terms related to evanescent operators.
Forgetting about them would produce a wrong result.

\subsection{Two-loop anomalous dimensions}
\label{S:u2}

\begin{figure}[ht]
\begin{center}
\begin{picture}(77.8,47.2)
\put(11.3,38.9){\makebox(0,0){\includegraphics{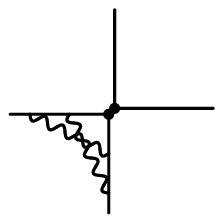}}}
\put(38.9,38.9){\makebox(0,0){\includegraphics{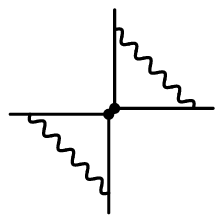}}}
\put(11.3,11.3){\makebox(0,0){\includegraphics{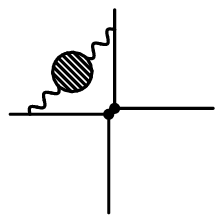}}}
\put(38.9,11.3){\makebox(0,0){\includegraphics{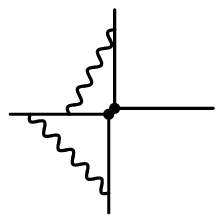}}}
\put(66.5,11.3){\makebox(0,0){\includegraphics{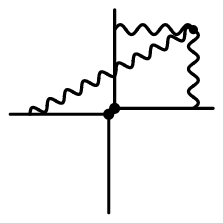}}}
\end{picture}
\end{center}
\caption{Some two-loop diagrams for the vertex functions of $O_{1,2}$.}
\label{F:V2}
\end{figure}

Some typical diagrams for the calculation of the two-loop anomalous dimension matrix
are shown in Fig.~\ref{F:V2}.
We need only UV $1/\varepsilon$ divergences;
the most efficient way to calculate them is to set all external momenta to 0,
and to insert a small mass $m$ into all denominators as an IR regulator~\cite{CMM:98b,CMM:98}.
This is similar to what we did at one loop.

\begin{figure}[ht]
\begin{center}
\begin{picture}(20,26)
\put(10,13){\makebox(0,0){\includegraphics{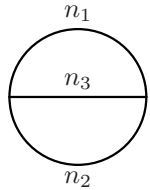}}}
\put(10,23){\makebox(0,0)[b]{$n_1$}}
\put(10,14){\makebox(0,0)[b]{$n_3$}}
\put(10,3){\makebox(0,0)[t]{$n_2$}}
\end{picture}
\end{center}
\caption{The two-loop massive vacuum integral.}
\label{F:int2}
\end{figure}

Then all diagrams reduce to the Euclidean scalar integrals (Fig.~\ref{F:int2})
\begin{equation}
\frac{1}{\pi^d}
\int \frac{d^d k_1\,d^d k_2}%
{(k_1^2 + m^2)^{n_1} (k_2^2 + m^2)^{n_2} ((k_1-k_2)^2 + m^2)^{n_3}}
= I_{n_1 n_2 n_3} m^{2(d-n_1-n_2-n_3)}\,.
\label{u2:I}
\end{equation}
If one of the indices is $\le0$,
it reduces to a trivial product of one-loop integrals.
When all the indices are $>0$,
we can use integration by parts~\cite{DT:93}:
\begin{equation}
\left[ d - 3 n_1 + 3 n_1 \1+
+ n_2 \2+ (\3- - \1-) + n_3 \3+ (\2- - \1-) \right] I = 0\,.
\end{equation}
This relation, together with symmetric ones,
reduces any $I_{n_1 n_2 n_3}$ to trivial cases
and a single non-trivial master integral $I_{111}$.

It can be found using Mellin--Barnes representation
\begin{equation}
\begin{picture}(12,5)
\put(6,1){\makebox(0,0){\includegraphics{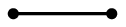}}}
\put(6,2){\makebox(0,0)[b]{$n$}}
\end{picture} =
\frac{1}{\Gamma(n)} \frac{1}{2\pi i} \int_{-i\infty}^{+i\infty} dz\,
\Gamma(-z) \Gamma(n+z) m^{2z}
\begin{picture}(12,5)
\put(6,1){\makebox(0,0){\includegraphics{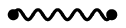}}}
\put(6,2){\makebox(0,0)[b]{$n+z$}}
\end{picture}
\,.
\label{u2:MB}
\end{equation}
Substituting
\begin{equation}
\raisebox{-11.75mm}{\begin{picture}(20,26)
\put(10,13){\makebox(0,0){\includegraphics{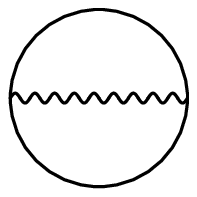}}}
\put(10,23){\makebox(0,0)[b]{$n_1$}}
\put(10,14){\makebox(0,0)[b]{$n_3$}}
\put(10,3){\makebox(0,0)[t]{$n_2$}}
\end{picture}} =
\frac{\Gamma\left(\frac{d}{2}-n_3\right)
\Gamma\left(n_1+n_3-\frac{d}{2}\right)\Gamma\left(n_2+n_3-\frac{d}{2}\right)
\Gamma(n_1+n_2+n_3-d)}%
{\Gamma\left(\frac{d}{2}\right)\Gamma(n_1)\Gamma(n_2)\Gamma(n_1+n_2+2n_3-d)}\,,
\label{u2:V}
\end{equation}
we obtain~\cite{DT:93}
\begin{equation}
\begin{split}
&I_{n_1 n_2 n_3} =
\frac{1}{\Gamma\left(\frac{d}{2}\right) \Gamma(n_1) \Gamma(n_2) \Gamma(n_3)}
\frac{1}{2\pi i} \int_{-i\infty}^{+i\infty} dz\,
\Gamma(-z) \Gamma\left(\tfrac{d}{2}-n_3-z\right)\\
&\frac{\Gamma(n_3+z) \Gamma\left(n_1+n_3-\frac{d}{2}+z\right) \Gamma\left(n_2+n_3-\frac{d}{2}+z\right)
\Gamma(n_1+n_2+n_3-d+z)}{\Gamma(n_1+n_2+2n_3-d+2z)}\,.
\end{split}
\label{u2:Andrei}
\end{equation}
We can close the integration contour to the right.
there are two series of right poles, $z=n$ and $z=n+\frac{d}{2}-n_3$,
producing two hypergeometric series.
In particular, the master integral is~\cite{DT:93}
\begin{equation}
I_{111} = \frac{\Gamma^2(\varepsilon)}{1-\varepsilon} \left[
\F{2}{1}{1,\varepsilon\\\frac{3}{2}}{\frac{1}{4}}
+ \frac{1}{1-2\varepsilon}
\F{2}{1}{1,-1+2\varepsilon\\\frac{1}{2}+\varepsilon}{\frac{1}{4}}
\right]\,.
\end{equation}
This exact result can be expanded in $\varepsilon$.

Note that $\gamma_5$ is not used in the calculation~\cite{CMM:98}:
it is hidden in the index $L$ of the external fermion wave functions.
These wave functions determine which $\gamma$-matrix structures
vanish at $d=4$, and hence which operators are evanescent.

\section{$b \to s$}
\label{S:s}

In this section we shall consider processes
in which the number of $b$ quarks reduce by 1,
the number of $s$ quarks increases by 1,
and the other flavor numbers don't change.
We shall consider the lowest order in electroweak interactions,
but taking into account QCD corrections.
The process $b\to s\gamma$ requires an additional factor $e$.
We shall not discuss it here.
Several additional operators appear in the effective Lagrangian
at the next order in electroweak interaction,
but it is not difficult to extend the methods discussed here
to $b\to s\gamma$.

\begin{figure}[ht]
\begin{center}
\begin{picture}(38,29)
\put(19,14.5){\makebox(0,0){\includegraphics{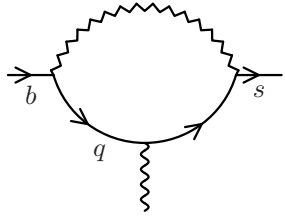}}}
\put(4,17.5){\makebox(0,0)[t]{$b$}}
\put(34,17.5){\makebox(0,0)[t]{$s$}}
\put(13,8.5){\makebox(0,0){$q$}}
\end{picture}
\end{center}
\caption{The $bsg$ vertex.}
\label{F:bsg}
\end{figure}

At first sight one might think that the diagram in Fig.~\ref{F:bsg}
can produce the operator
\[
g\,\bar{s}_L G^a_{\mu\nu} t^a \sigma^{\mu\nu} b
\]
of dimension 5.
But here $b$ must be $b_R$,
otherwise the operator vanishes at $d=4$;
and this is impossible at $m_b=0$.
Therefore in fact the operator
\begin{equation}
O_g = g m_b \bar{s}_L G^a_{\mu\nu} t^a \sigma^{\mu\nu} b_R
\label{s:Og}
\end{equation}
of dimension 6 is produced.
It is called the gluon dipole operator;
it is a mixture of magnetic and electric dipole interactions.

The coefficient of this operator in the effective Lagrangian is
\begin{equation}
\frac{g_2^2}{M_W^2} \sum_{q=u,c,t} V_{qb} V^*_{qs} E(x_q)\,,
\label{s:Cg1}
\end{equation}
where
\begin{equation}
x_q = \frac{m_q^2}{M_W^2}\,,
\label{s:xq}
\end{equation}
and the function $E(x_q)$ can be easily calculated
from the vacuum integral of Fig.~\ref{F:bsg}
which depends on two masses, $M_W$ and $m_q$.
But
\begin{equation}
\sum_{q=u,c,t} V_{qb} V^*_{qs} = 0
\label{s:V}
\end{equation}
due to unitarity of the matrix $V$.
Therefore we can rewrite~(\ref{s:Cg1}) as
\[
\frac{g_2^2}{M_W^2} \sum_{q=u,c,t} V_{qb} V^*_{qs} \left[E(x_q)-E(0)\right]\,.
\]
The only $x_q$ substantially different from 0 is $x_t$;
therefore, the coefficient of the dipole operator~(\ref{s:Og})
in the effective Lagrangian is
\begin{equation}
\frac{g_2^2}{M_W^2}V_{tb} V^*_{ts} \left[E(x_t)-E(0)\right]\,.
\label{s:Cg}
\end{equation}

In order to obtain a non-vanishing contribution
from the on-shell diagram in Fig.~\ref{F:bsg}
we expanded in in $m_b$ up to the linear term.
Alternatively, we can expand it in the gluon momentum $q$
up to the linear term, and obtain a non-vanishing operator
\begin{equation}
g\,\bar{s}_L D^\nu G^a_{\mu\nu} t^a \gamma^\mu b_L\,.
\label{s:OpG}
\end{equation}
Due to the QCD equation of motion,
\begin{equation}
D^\nu G^a_{\mu\nu} = g \sum_q \bar{q} t^a \gamma_\mu q\,.
\label{s:EOM}
\end{equation}
Therefore we can rewrite the operator~(\ref{s:OpG}) as
\begin{equation}
O_p = g^2 \left(\bar{s}_L t^a \gamma^\alpha b_L\right)
\sum_q \left(\bar{q} t^a \gamma_\alpha q\right)\,,
\label{s:Op}
\end{equation}
up to an EOM-vanishing operator.
On-shell matrix elements of EOM-vanishing operators vanish;
we can safely omit them from the effective Lagrangian,
which is constructed to reproduce the correct $S$-matrix.
The operator $O_p$ is called penguin%
\footnote{In 1977 John Ellis made a bet with Melissa Franklin at a bar:
if he loses a game of darts, he has to use the word ``penguin'' in his next paper.
He lost, and has drawn the diagram in Fig.~\ref{F:bsg} in a penguin-like shape.}.
Its coefficient in the Lagrangian is given by a formula
similar to~(\ref{s:Cg}).

\begin{figure}[ht]
\begin{center}
\begin{picture}(84,40)
\put(18.5,21.5){\makebox(0,0){\includegraphics{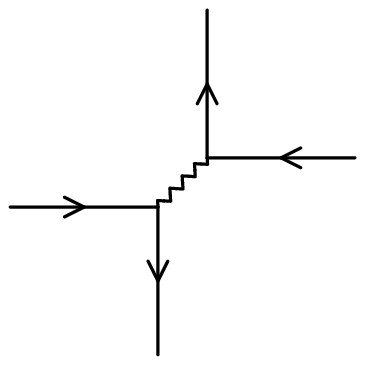}}}
\put(8.5,21.5){\makebox(0,0){$b$}}
\put(18.5,11.5){\makebox(0,0){$c$}}
\put(18.5,31.5){\makebox(0,0){$s$}}
\put(28.5,21.5){\makebox(0,0){$c$}}
\put(16.5,23.5){\makebox(0,0){$W$}}
\put(18.5,0){\makebox(0,0)[b]{a}}
\put(65.5,21.5){\makebox(0,0){\includegraphics{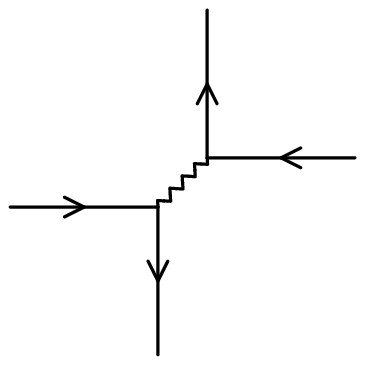}}}
\put(55.5,21.5){\makebox(0,0){$b$}}
\put(65.5,11.5){\makebox(0,0){$u$}}
\put(65.5,31.5){\makebox(0,0){$s$}}
\put(75.5,21.5){\makebox(0,0){$u$}}
\put(63.5,23.5){\makebox(0,0){$W$}}
\put(65.5,0){\makebox(0,0)[b]{b}}
\end{picture}
\end{center}
\caption{$b \to c \bar{c} s$ (a) and $b \to u \bar{u} s$ (b).}
\label{F:ccs}
\end{figure}

Of course, there is also the operator $O_{c1}$ (Fig.~\ref{F:ccs}a);
we need a set of operators closed under renormalization,
and hence have to include also $O_{c2}$:
\begin{equation}
O_{c1} = (\bar{c}_{Li} \gamma^\alpha b_L^i)\,
(\bar{s}_{Lj} \gamma_\alpha c_L^j)\,,\qquad
O_{c2} = (\bar{c}_{Li} \gamma^\alpha b_L^j)\,
(\bar{s}_{Lj} \gamma_\alpha c_L^i)\,.
\label{s:Oc}
\end{equation}
The similar operators
\begin{equation}
O_{u1} = (\bar{u}_{Li} \gamma^\alpha b_L^i)\,
(\bar{s}_{Lj} \gamma_\alpha u_L^j)\,,\qquad
O_{u2} = (\bar{u}_{Li} \gamma^\alpha b_L^j)\,
(\bar{s}_{Lj} \gamma_\alpha u_L^i)
\label{s:Ou}
\end{equation}
have CKM-suppressed coefficients.

Similarly, we should take into account not just one penguin operator $O_p$~(\ref{s:Op}),
but both color structures:
\begin{equation}
O_{p1} = (\bar{s}_{Li} \gamma^\alpha b_L^i)\,
\sum_q (\bar{q}_j \gamma_\alpha q^j)\,,\qquad
O_{p2} = (\bar{s}_{Li} \gamma^\alpha b_L^j)\,
\sum_q (\bar{q}_j \gamma_\alpha q^i)
\label{s:Op1}
\end{equation}
($O_p = T_F g^2 (O_{p2}-O_{p1}/N_c)$).
Unlike the operators $O_{1,2}$~(\ref{u:O12}) (or $O_{c1,2}$~(\ref{s:Oc})),
the penguin operators contain full quark fields $q$ in $\Sigma_q$,
not just their $L$ components.
Therefore the operators
\begin{equation}
O_{p3} = (\bar{s}_{Li} \gamma^\alpha \gamma^\beta \gamma^\gamma b_L^i)\,
\sum_q (\bar{q}_j \gamma_\gamma \gamma_\beta \gamma_\alpha q^j)\,,\qquad
O_{p4} = (\bar{s}_{Li} \gamma^\alpha \gamma^\beta \gamma^\gamma b_L^j)\,
\sum_q (\bar{q}_j \gamma_\gamma \gamma_\beta \gamma_\alpha q^i)
\label{s:Op3}
\end{equation}
with 3 $\gamma$-matrices don't reduce to~(\ref{s:Op1}) plus evanescent operators,
and should be included in our full set of operators.
On the other hand, the operators with 5 $\gamma$ matrices
do reduce to~(\ref{s:Op1}), (\ref{s:Op3}) plus evanescent ones.

Thus we arrive at the effective Lagrangian for $b\to s$ processes:
\begin{equation}
\begin{split}
L = \frac{g_2^2}{2 M_W^2} \biggl[&
V^*_{cs} V_{cb} \left( c_{c1} O_{c1} + c_{c2} O_{c2} \right)
+ V^*_{us} V_{ub} \left( c_{u1} O_{u1} + c_{u2} O_{u2} \right)\\
&{} + V^*_{ts} V_{tb} \biggl( c_g O_g + \sum_{i=1}^4 c_{pi} O_{pi} \biggr)
\biggr]\,.
\end{split}
\label{s:L}
\end{equation}
The Wilson coefficients at $\mu=M_W$ are obtained by matching:
$c_{c1}$, $c_{u1}$, $c_g$ are $1+\mathcal{O}(\alpha_s)$;
$c_{c2}$, $c_{u2}$, $c_{pi}$ are $\mathcal{O}(\alpha_s)$.
In order to find them at a low $\mu\sim m_b$,
we need to solve the RG equations~(\ref{u:RG1}),
and hence we need the anomalous dimension matrix of these operators.

Processes like $b\to s\gamma$
involve an extra electromagnetic interaction, and require some additional operators.
There is the photon dipole operator $O_\gamma$ similar to the gluon one $O_g$~(\ref{s:Og},
and photon penguin operators similar to~(\ref{s:Op1}), (\ref{s:Op3}).

\section{$B^0 \leftrightarrow \bar{B}^0$}
\label{S:osc}

\begin{figure}[ht]
\begin{center}
\begin{picture}(73,24)
\put(17,12){\makebox(0,0){\includegraphics{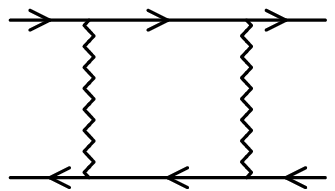}}}
\put(5,22){\makebox(0,0){$\vphantom{q}b$}}
\put(17,22){\makebox(0,0){$\vphantom{b}q$}}
\put(29,22){\makebox(0,0){$\vphantom{q}d$}}
\put(5,2){\makebox(0,0){$\vphantom{q}d$}}
\put(17,2){\makebox(0,0){$\vphantom{b}q'$}}
\put(29,2){\makebox(0,0){$\vphantom{q}b$}}
\put(56,12){\makebox(0,0){\includegraphics{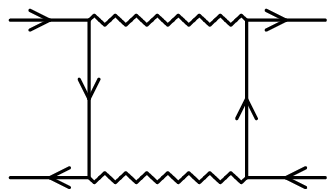}}}
\put(44,22){\makebox(0,0){$\vphantom{q}b$}}
\put(46,12){\makebox(0,0){$\vphantom{q'}q$}}
\put(44,2){\makebox(0,0){$\vphantom{q}d$}}
\put(68,22){\makebox(0,0){$\vphantom{q}d$}}
\put(66,12){\makebox(0,0){$q'$}}
\put(68,2){\makebox(0,0){$\vphantom{q}b$}}
\end{picture}
\end{center}
\caption{Diagrams of $b\bar{d}\leftrightarrow d\bar{b}$ transitions.}
\label{F:box}
\end{figure}

Finally, we shall briefly discuss a process which at the order $g_2^4$:
$B^0 \leftrightarrow \bar{B}^0$ oscillations (Fig.~\ref{F:box}).
They are described by the effective Lagrangian
\begin{equation}
L = \frac{g_2^4}{512 \pi^2 M_W^2} c O\,,\qquad
O = (\bar{d}_L \gamma^\alpha b_L)\,
(\bar{d}_L \gamma^\alpha b_L)\,.
\label{osc:L}
\end{equation}
The Wilson coefficient is given by
\begin{equation}
c = \sum_{q,q'=u,c,t} V^*_{qb} V_{qd} V^*_{q'b} V_{q'd} S(x_q,x_{q'})\,,
\label{osc:c1}
\end{equation}
where $S(x_q,x_{q'})=S(x_{q'},x_q)$ is given by the one-loop vacuum integrals (Fig.~\ref{F:box})
with three masses: $M_W$, $m_q$, $m_{q'}$
($x_q$ is defined by~(\ref{s:xq})).
Due to~(\ref{s:V}),
\[
c = \sum_{q,q'=u,c,t} V^*_{qb} V_{qd} V^*_{q'b} V_{q'd} \left[S(x_q,x_{q'}) - S(x_q,0)\right]
= V^*_{tb} V_{td} \sum_{q=u,c,t} V^*_{qb} V_{qd} \left[S(x_q,x_t) - S(x_q,0)\right]\,,
\]
because only $x_t$ substantially differs from 0.
Finally,
\begin{equation}
\begin{split}
c &= V^*_{tb} V_{td} \sum_{q=u,c,t} V^*_{qb} V_{qd} \left[S(x_q,x_t) - S(x_q,0) - S(0,x_t) + S(0,0)\right]\\
&= \left(V^*_{tb} V_{td}\right)^2 \left[S(x_t,x_t) - 2 S(x_t,0) + S(0,0)\right]\,.
\end{split}
\label{osc:c2}
\end{equation}

We don't need to calculate the one-loop anomalous dimension of the operator $O$~(\ref{osc:L})
because it has been already done in Sect.~\ref{S:u1}:
\begin{equation}
\gamma_0 = \lambda_+ = 12 T_F \left( 1 - \frac{1}{N_c} \right)\,,
\label{osc:gamma}
\end{equation}
see~(\ref{u:vpm})
(the operator similar to $O_-$ is zero).

\section{Conclusion}
\label{S:Conc}

I am grateful to the organizers of the Dubna school on heavy quark physics
for inviting me.
The work was partially supported by RFBR (grant 12-02-00106-a)
and by Russian Ministry of Education and Science.
Writing of this text was done at the Universities of Mainz and Siegen
and Karlsruhe Institute of Technology;
I am grateful to M.~Neubert, T.~Mannel, and M.~Steinhauser
for their hospitality.

\begin{footnotesize}

\end{footnotesize}
\end{document}